# Visualization of Board of Director Connections for Analysis in Socially Responsible Investing

Alice Da Fonseca, PhD.
*College of Arts & Science*
*Tufts University*
Medford, USA
alice.mello@tufts.edu

Peter Lake
*College of Professional Studies*
*Northeastern University*
Boston, USA
lake.p@northeastern.edu

Ariana Barrenechea
*Data Science*
*Free Float LLC*
Connecticut, USA
ari@freefloat.llc

*Abstract*—This project is a collaboration between industry and academia to delve into Finance Social Networks, specifically the Board of Directors of public companies. Knowing the connections between Directors and Executives in different companies can generate powerful stories and meaningful insights on investments. A proof of concept in the form of a Data Visualization tool reveals its strength in investigating corporate governance and sustainability, as well as in the partnership between industry and academic institutions.

*Keywords—network data viz, executives' social network, corporate governance, board members, investing.*

I. INTRODUCTION

The impact of social networks in finance has been analyzed in the literature about Corporate Finance Policies and Social Networks [3], concluding that there is an increase in similar capital investments when social ties between the company directors and/or board members exist, and the strength of this connection might also be affected by regions or sectors. Fracassi's study takes into consideration the similarities and connections through education, profession, management styles, etc. among each pair of executives following a similar analysis done by Shue [5], Cohen [2], Hong [4], and Clement [1], among others. It suggests that social connections can have a causal effect on corporation policies. He also based his study on Network Analysis, suggesting that depending on how strategically companies are positioned in the network can influence their investment decisions.

With those studies and a series of practical cases in mind, Free Float LLC, our sponsor company, is exploring how the tenure of board members and/or directors can positively or negatively influence the companies' policies and investments. As their mission states on their website: 'Our focus is understanding the people behind companies through data and narratives. "The Market" tends to be the most important protagonist to every business story. We believe the market is a collection of people making decisions, and investors should measure those people first and the companies second. Investors haven't had the tool to do that until now.'

Free Float LLC requested a dashboard that could represent the connections between members of the Board of Directors from thousands of international companies. The purpose was to help investors navigate through the interconnectedness of board members to understand better the possible conflicts companies faced in meeting their environmental and social responsibilities. They believe that quantifying social dynamics inside a board room and visualizing those social connections along with a series of specific attributes of those directors and board members could provide insight on the people behind the companies and the direction of the company. Investors could then use this information to adjust their investing strategies to match their social responsibility goals.

The request became an experiential learning (XN) project for a Bachelor of Science in Analytics degree course on Data Visualization (ALY 3070 Communication and Visualization for Data Analytics) at Northeastern University. This paper details the project development and describes the tool that resulted. It also illustrates how powerful XN can be as a real-world learning experience in intermediate courses and the benefits for the company sponsoring the project.

The data our sponsor receives from their data provider has more than 50 KPIs for almost 9,000 companies and over 80,000 current directors worldwide. When considering a nine-year period from 2015 to 2023, the list of individuals current and retired grows to 200,000. Each director might have several records of information, which generates over a million records for this time frame. Free Float updates their data once a month. Certain KPIs are stagnant (such as name, gender, etc.) but other metrics (such as board membership, market capitalization, board composition, etc.) change at least once a year for most companies.

Free Float LLC provided this information to the students in two data sets, which were a comprehensive snapshot of data available, but not fully complete: 'director_independence_factors.csv' (DIF) and 'board_connections_edges.csv' (BCE). DIF contained information on 67,515 unique directors and almost 9,000 companies. DIF consisted of factors such as gender, CEO status and time frame, Chairman status and timeframe, Founder status, and other variables that might indicate the director's influence on the company. In addition, this dataset included a proprietary influence metric (named INF) devised by Free Float LLC to indicate a director's influence on a company. This metric was given for 2018, 2019, 2020, 2021, and 2022 up to the date we received the dataset. Fig. 1 shows those datasets split into five tables for easier analysis (DirectorInfo, UniqueDirectorInfo, CompanyInformation, Country LU, and INF by Country All).

DIF and BCE files were imported into PowerBI and several modifications and aggregations were made to enable a comprehensive analysis:

- DIF was renamed to DirectorInfo, and several data types changed. One new calculated field was created to hold the average of the previous four years' INF for the director/company and a second was created to hold the market cap league description as the original data just had a numeric code.

- Because there were multiple records for directors (one for each company they were associated with) in the DIF there needed to be a lookup table (UniqueDirectorInfo) with a unique record for each director that contained the basic biographical information which could be linked to other tables. There were some directors with mismatched data (same Director ID but different name spellings, missing gender data, inconsistent CoServed and Activist fields) in the DIF so the extract from the DIF into the table was created by grouping on Director ID and taking the Max value of all the other fields. This was an arbitrary decision but because of the small number of records involved, it would not affect the analysis.

- Similarly, company information contained in the DIF was also duplicated, so a lookup table (CompanyInformation) was created. A small number of records had mismatched data for the same company (different values for MktCap, IndSec, or Country) so, again, the data was grouped on CompanyName, and the Max of the other fields was used.

- BCE was renamed to Connections. Each record contains the company ID and name, the Node1 director ID and name, the Node2 director ID and name, and the overlap time. Node1 and Node2 have no information value, in other words, a director could arbitrarily be either Node1 or Node2 for any specific connection.

- Because the DIF used ISO 3166 two-letter country codes a lookup table (Country LU) was created by downloading the ISO two-letter country code table.

- A table was created (INF_by_Country_All) to hold a subset of company data (INF, Market Cap, Sector, country) that had the INF_YR columns (inf_today, inf_2000, etc) transposed into rows.

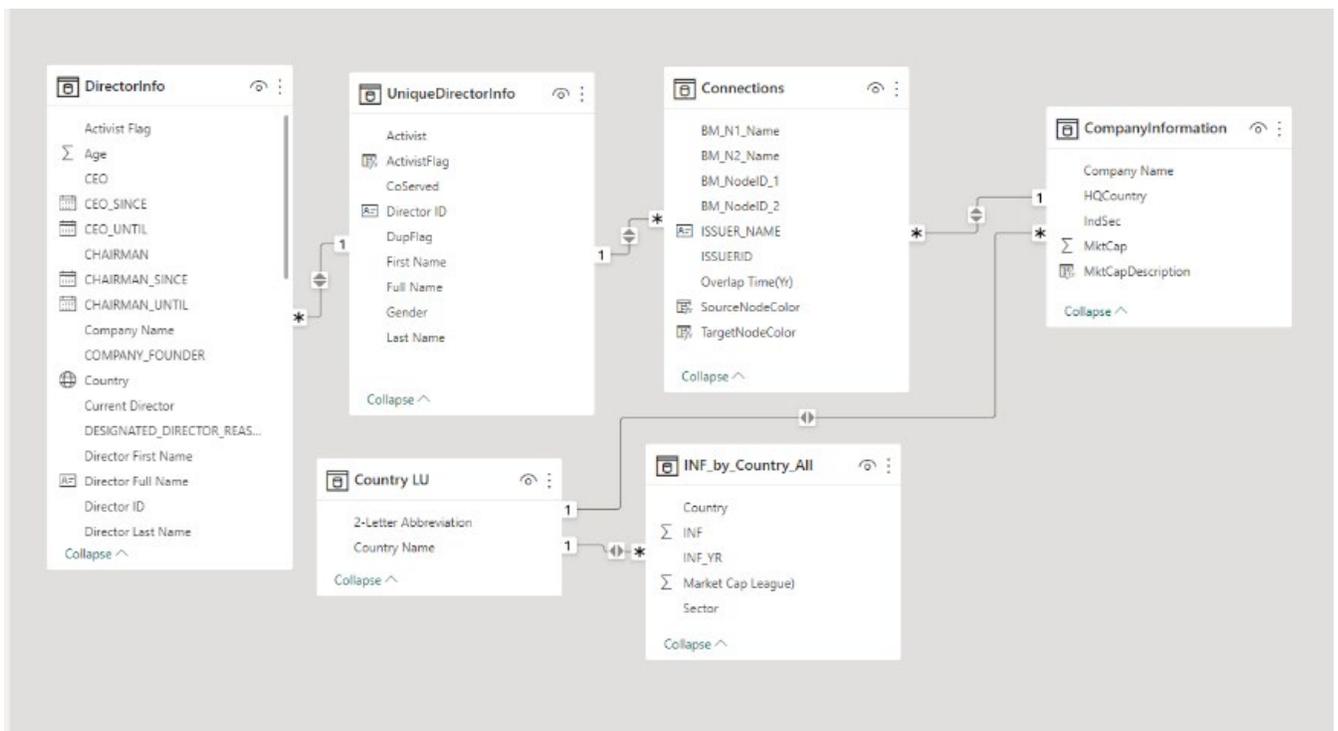

Fig. 1.  Data Mapping

| Original files | |
|---|---|
| director_independence_factors.csv (DIF) | 80,610 records |
| board_connections_edges.csv (BCE) | 1,164,584 records |
| **New tables** | | |
| *Name* | *Source* | *Size* |
| DirectorInfo | DIF | 80,610 records |
| UniqueDirectorInfo | DIF | 67,515 records |
| CompanyInformation | DIF | 8,939 records |
| INF_by_Country_All | DIF | 322,440 records |
| Connections | BCE | 1,164,584 records |
| Country LU | ISO 3166 | 64 records |

Table 1. Processed Data

The students were given the task of exploring the real finance data and finding a way to best map the network among directors, including as much relevant data as possible to give investors information that could be used in their investment decisions. The resulting tool should give the end users the ability to explore the data on their own using filtering and sorting. They were able to contact the industry sponsor with any questions about the data or to confirm any of their insights throughout the six weeks of the project. They used Microsoft Power BI to transform and analyze the data and tell their stories through visuals such as Network Navigator, Force-Directed Graph 2.02, and ArcGIS Maps for Power BI among others.

## II. AUDIENCE

ALY3070 is a course based on the concepts of Storytelling with Data. This challenge posed by Free Float LLC was a great chance for the students to explore quantitative and qualitative data while learning a diverse range of visualization in Power BI, the tool used by the sponsor. They needed to create visualizations that could best tell the stories to investors looking for connections between companies or trying to identify individuals who had some degree of influence within a company or industry. They could also create stories of interest to someone looking at social science factors such as gender, age, or geographic base and their relation to influence or connectivity.

Some students crafted their stories for other kinds of audiences as well, for example, anyone looking for a job or exploring the gender disparities between the company directors by country. There was latitude within the project for the students to generate business questions different from the one the company sponsor requested. This gave them a diverse audience to tell their stories to and allowed them to create a variety of dashboards.

As an example, using this tool, Free Float found the gender power disparity in the United States to be less than in countries that have "board quotas" such as Norway. In Norway, even though public boards are required to have gender diversity at the board level and companies meet this criterion, the female board members typically are not put in positions of power (i.e. CEO, Committee Chair/Member, or Lead Director) therefore they tend to have less influence overall. This surprising fact was of special interest to the Northeast Investors' Diversity Initiative, made up of six state treasuries including CT and MA. This initiative is "a coalition of institutional investors committed to increasing gender, racial, and ethnic diversity on corporate boards to maximize returns and safeguard shareholder value." Since gender diversity is a specific social responsibility goal of this coalition, they can use this information to adjust their investing or demand change - either through engagement or through a formal process of shareholder proposals.

## III. THE DATA VIZ APPLICATION

The figures below capture one dashboard at a time along with their filters. The stories they tell start by showing the node connections (Figs. 2 - 4) and how one could view them filtered by Sector, Country, and Market Cap League (Small=<2B, Medium=2-10B, Large=10-200B, Mega=>200B). The goal is to allow comparison within and across these categories.

Fig. 2 shows an example of these connections. Other information existing in this visual includes overlap information, shown by the size of the nodes (blue and green circles), where larger circles indicate greater total overlap, and connections (gray lines) where thicker lines indicate greater average overlap.

This extra information allows one to make judgments about a director's influence. For example, a director with a longer total overlap (larger circle) but a smaller average overlap (thin connecting line) would mean they had many connections of shorter duration. One could infer they would have less influence within each company due to weaker social connections than a director with less total overlap but a larger average overlap (meaning they have spent more time with the directors they are connected to). Of course, identifying directors with both longer total and average overlap would be useful in identifying the 'linchpin' directors. These would be the directors that you would want to know more about so you could find out their priorities and how they would be likely to influence the company. From that information you would be able to make a more informed investment decision.

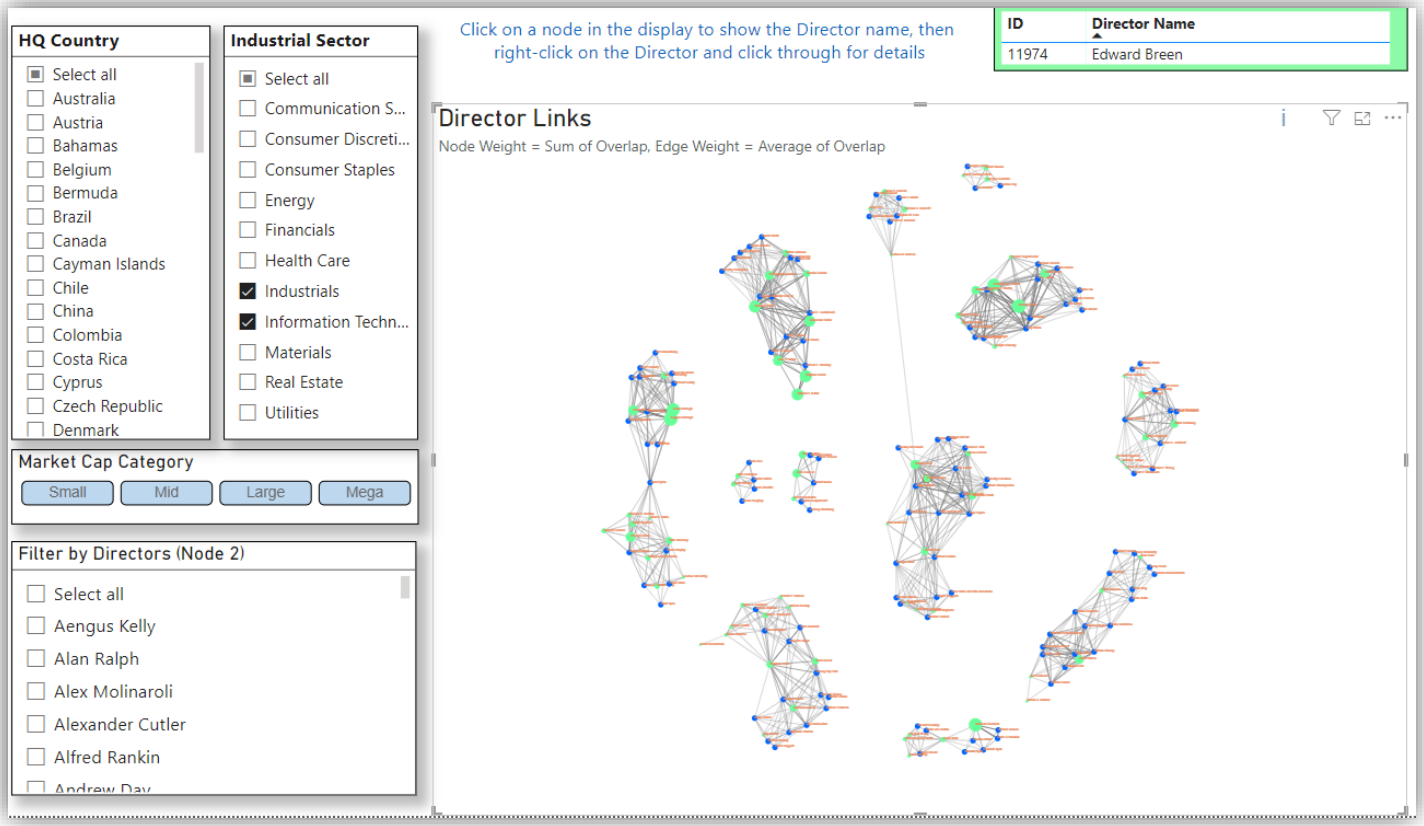

Fig. 2.    *Network of Directors*

Several things were discovered while creating and exploring this visual. First, structurally, it was important to set up filters to get a fairly small subset of data otherwise the graphic could not show all the connections, and the ones it did show were so densely packed that they didn't allow easy analysis. Fig. 3 demonstrates the limited usefulness of not having any filters selected.

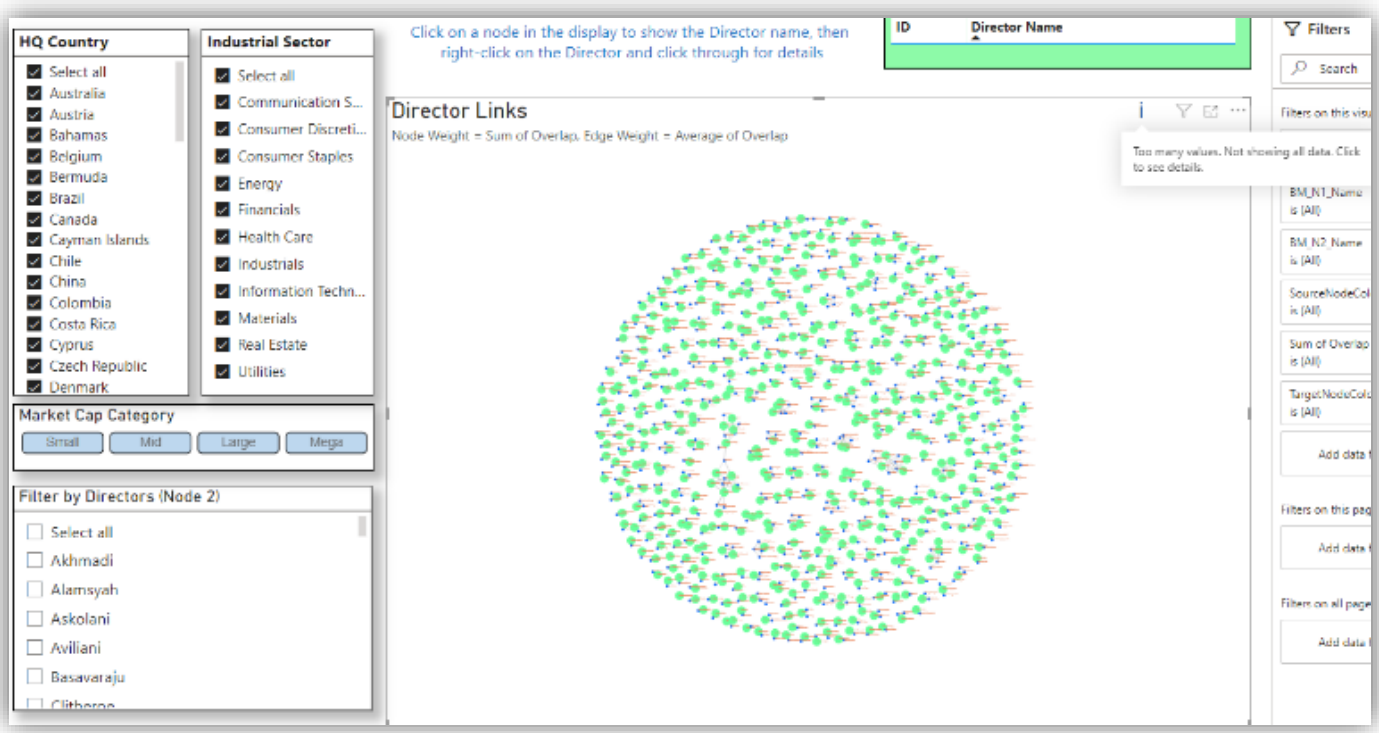

Fig. 3. *Network of Directors w/o Filter*

From a data standpoint, it was noted that there was much greater clustering within Industrial Sectors than across. This was especially prevalent in smaller market-cap companies. Large and Mega cap companies tended to have many more directors that crossed sectors. Fig. 4 shows a comparison between Mega Cap and Small Cap. Small Cap has many clusters with five or fewer connections, whereas Mega Cap shows a much smaller number of clusters but virtually all have more than ten connections. A possible conclusion from this would be less effective oversight in these larger companies (the Board has little to no experience in that industry), although that would need to be explored further.

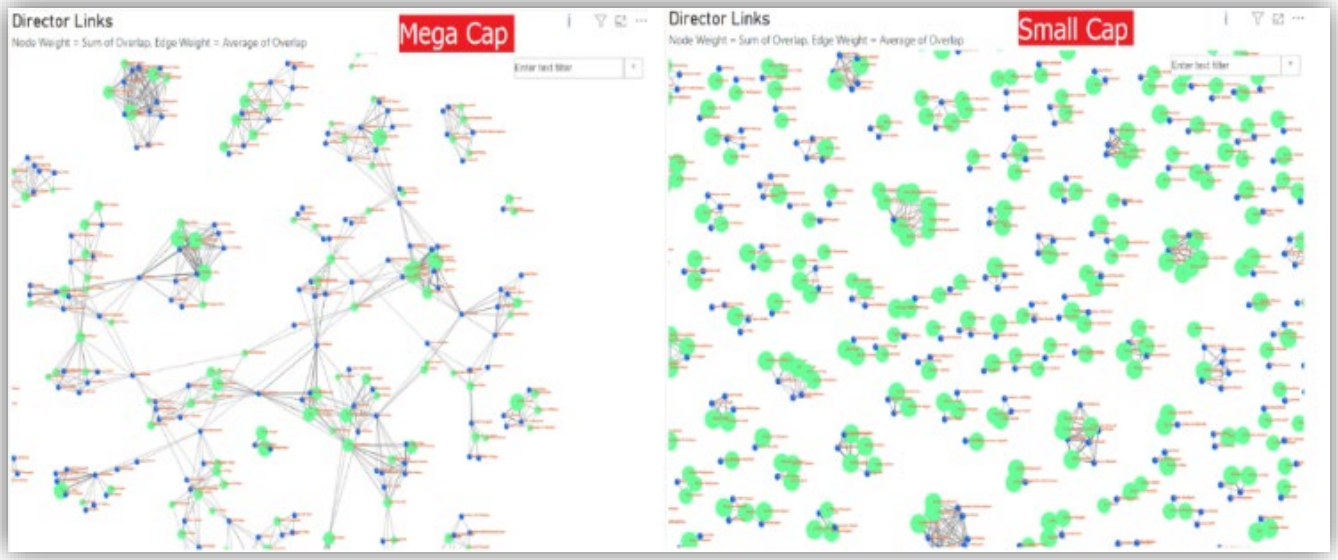

Fig. 4. *Small Cap vs Mega Cap Connections*

The user can see other connection information that might be hidden by the filtering selections on the Network of Directors (Fig.2) using the click-through feature. This is accessed by right-clicking on the selected Director's name in the top-right and clicking on Drill-Through. For example, Thomas Breen is selected, and the Director Detail Viz (Fig. 5a) shows that, in addition to the connections shown on the Network Viz, he is also on the Board of Directors of two other companies. These don't show on the visual because they are in different industries and/or in different countries.

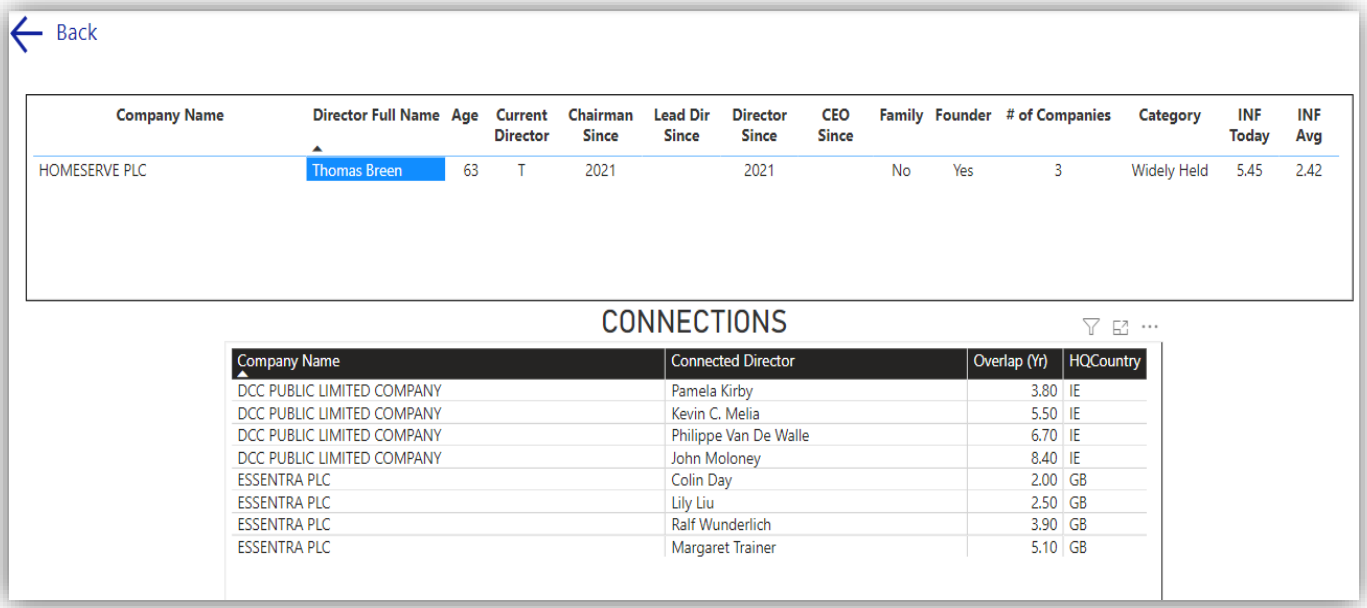

Fig. 5a. *Director Detail Showing Multiple Companies*

Fig. 5b illustrates another useful function of the Director Detail Viz. It is valuable in evaluating the reasons for the strength of the director's influence. In this case, it is shown that the director is also the Chairman and CEO. It also indicates that the company is considered Controlled (one shareholder controls the majority of the Board votes) and the assumption would be this shareholder is the one who controls it. These factors are reflected in his extraordinary INF score. He would be a good target if one was seeking to understand the company's direction or identify risk related to people in leadership positions.

Risk through the people in leadership lens is the next frontier of corporate governance, as evidenced most recently by TKO Group Holdings - the public entity after the merger of World Wrestling Entertainment (WWE) and Ultimate Fighting Championship. In a quarterly SEC filing, TKO noted of WWE founder and CEO Vince McMahon: "Mr. McMahon's membership on our Board could expose us to negative publicity and/or have other adverse financial and operational impacts on our business. His membership also may result in additional scrutiny or otherwise exacerbate the other risks described herein. Any of these outcomes could directly or indirectly have adverse financial and operational impacts on our business." [6].

| Company Name | Director Full Name | Age | Current Director | Chairman Since | Lead Dir Since | Director Since | CEO Since | Family | Founder | # of Companies | Category | INF Today | INF Avg |
|---|---|---|---|---|---|---|---|---|---|---|---|---|---|
| BERKSHIRE HATHAWAY INC. | Warren Buffett | 91 | T | 1970 | | 1965 | 1970 | Yes | No | 4 | Controlled | 64.13 | 64.39 |

CONNECTIONS

| Company Name | Connected Director | Overlap (Yr) | HQCountry |
|---|---|---|---|
| THE COCA-COLA COMPANY | Cathleen Black | 12.60 | US |
| THE COCA-COLA COMPANY | Ronald Allen | 15.30 | US |
| THE COCA-COLA COMPANY | Herbert Allen | 17.30 | US |

Fig. 5b. *Director Detail Showing High Influence*

The Company Connections Db (Fig. 6) shows a different node-link diagram, this time highlighting the companies involved in the connections. Each company is indicated by the same color links. Companies that are connected by many directors might be a red flag for conflicts of interest. This visual allows easy identification of that scenario and could serve for investment decisions. As Free Float explored this visual tool after its release, they found that Caterpillar, Boeing, and Marriott had at least three directors serving at these companies concurrently. This type of proven "friendship" at the board level is a group-think risk that investors want to identify. This visual also allows one to drill through to more specific information on the directors within a single company (Fig. 7), similar to the Director Detail visual.

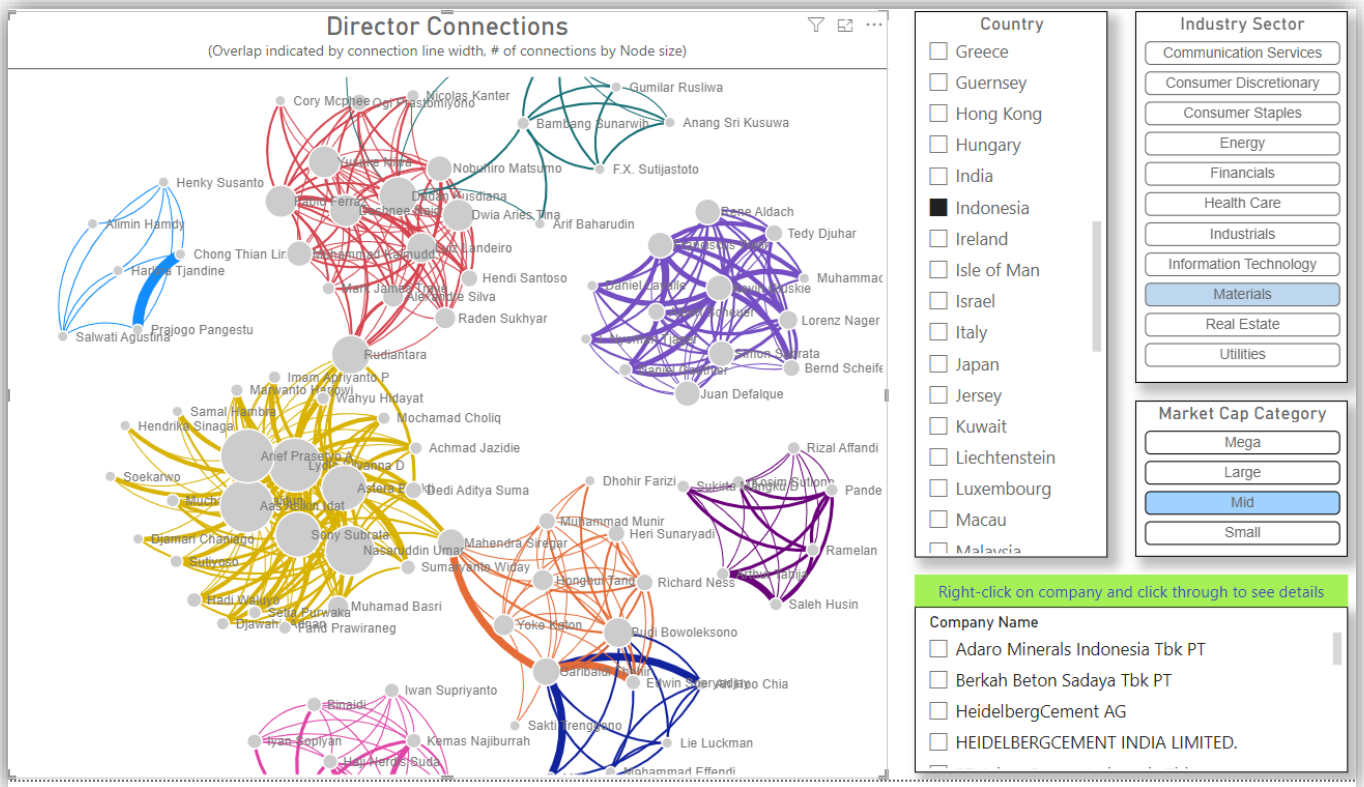

Fig. 6. *Company Connections (Alternative Network of Connections)*

Fig 7. *Company Directors*

The last three dashboards are all different views of the social science factors and their effect on influence, using the INF factor. There is a geographical distribution map (Fig. 8, 9), a Tenure (time on a board) summary (Fig. 10), and detail dashboard (Fig. 11) that shows the effect of Gender and Family Firm on Tenure and how this breaks down across countries and age.

Fig. 8 allows the user to see aggregated INF data by country, both discrete numbers and the trend. The summary data includes all countries by default, but by clicking on any country (green circle) the visuals update to show just that country's data. Fig. 9 shows South Korea isolated.

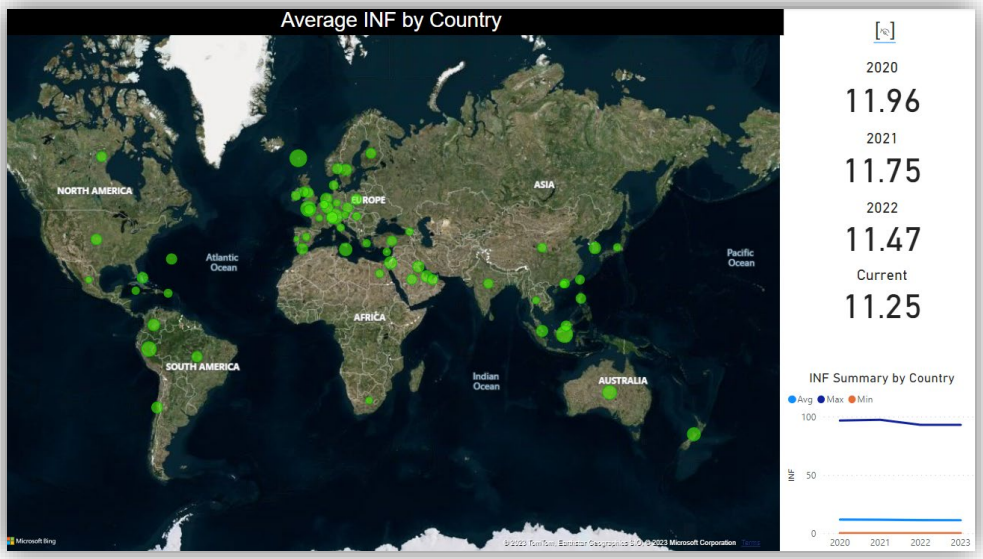

Fig. 8. *Influencers by Country*

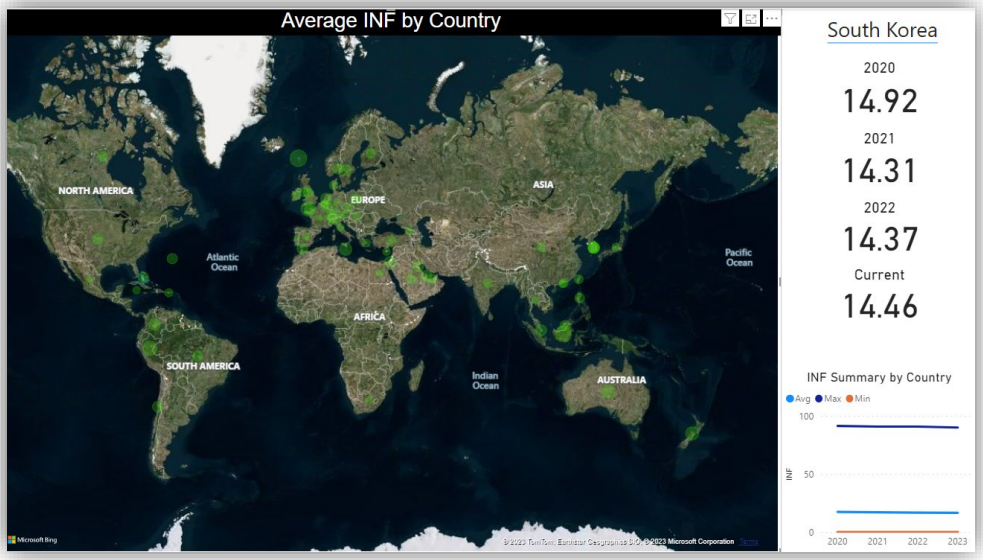

Fig. 9. *Influencers Filtered by Country*

One thing that was strikingly apparent from these visuals was the lack of any representation from African companies. From a social science standpoint, this would be strong evidence that African nations are still lagging in self-determination, as their economies are reliant on countries that have headquarters elsewhere.

The Executive Tenure visual was an exploration of how different factors affect Tenure (time on the board). Using the filters on this dashboard (Fig.10) the user can make comparisons on the effects of Country, Gender, and whether the firm is run by family. The average tenure across all industries, broken down by Market Cap League, is given at the top, and the bar graphs show the distribution across industrial sectors within Market Cap League. Right-clicking allows the user to drill through to more discrete details for a particular sector within a Market Cap League (Fig. 11).

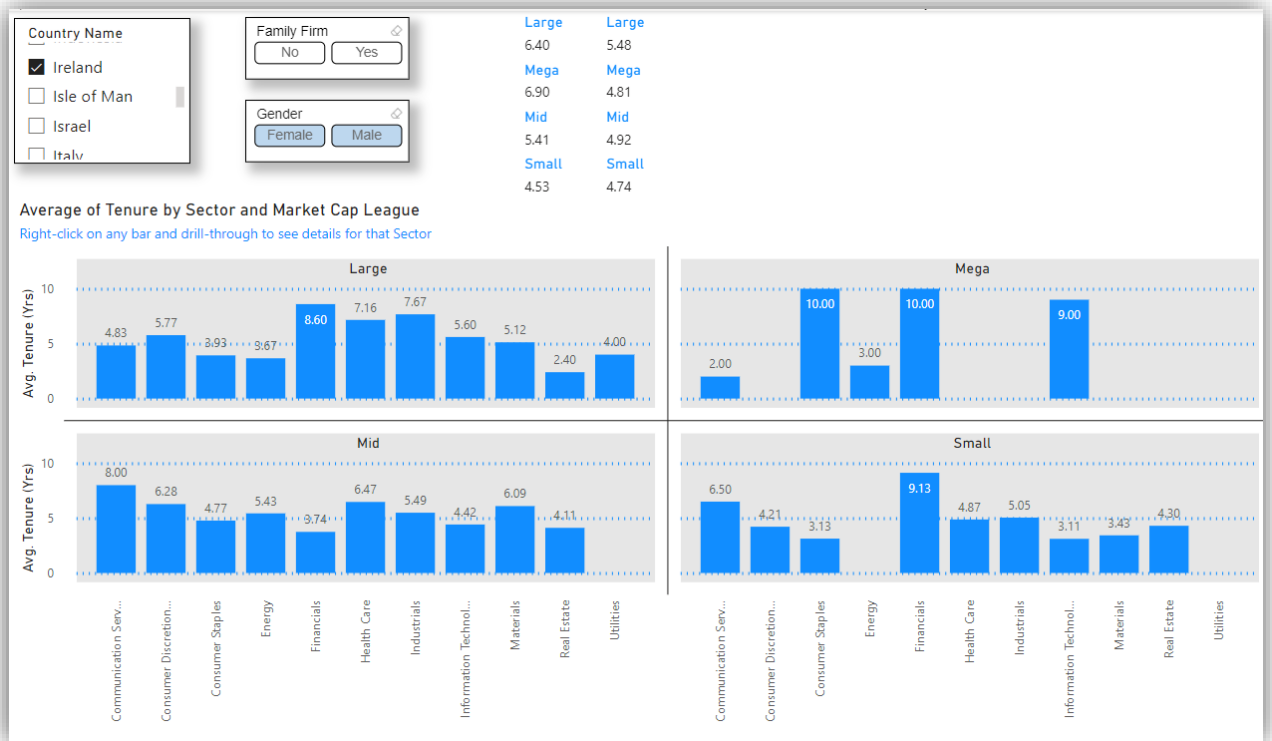

Fig. 10. *Executive Tenure*

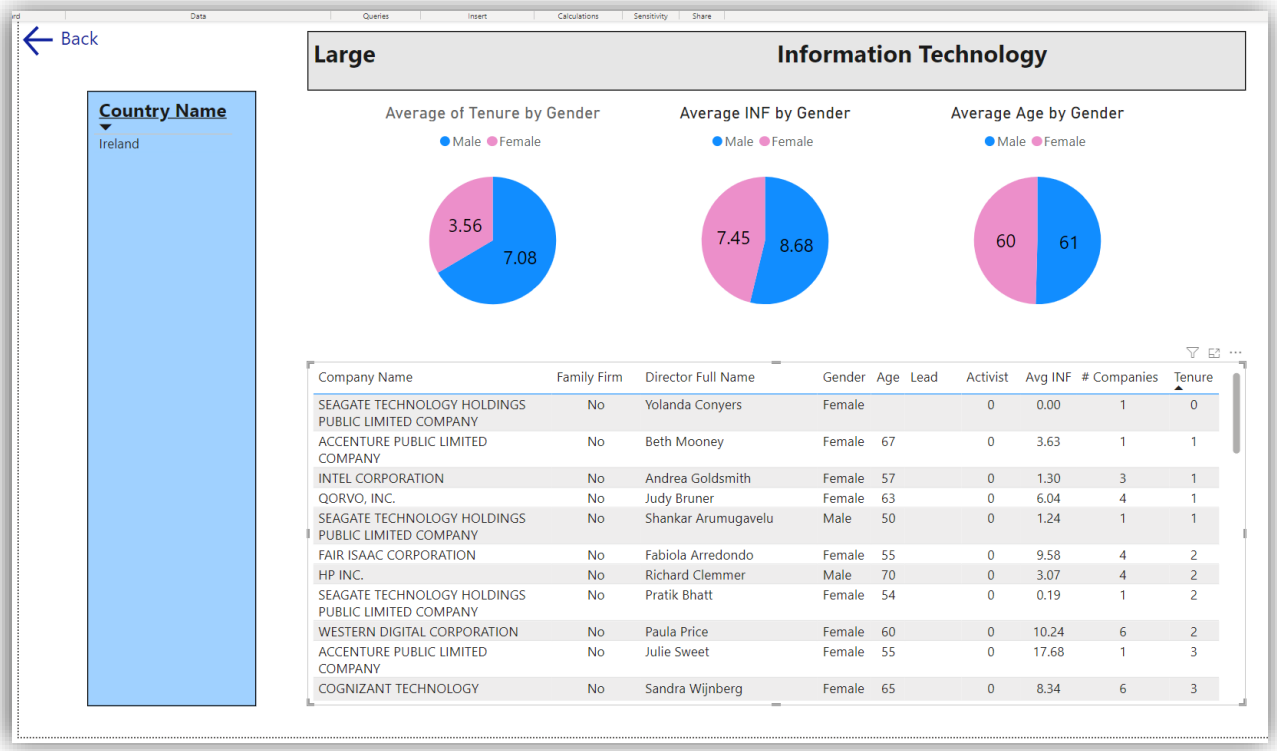

Fig. 11. *Detail on Director Tenure*

One of the interesting things that Free Float compared while using this tool was the tenure of the board of directors of Southwest Airlines and the average tenure of a non-family US firm of the same league. Southwest was in the news for several days as thousands of flights were disrupted and grounded over the holiday season in December 2022. When comparing the tenure of their board to the average, it was noted that Southwest directors doubled the average tenure length. Their age was also well above the average age of directors. This type of insight helps to tell the story of Southwest from the people perspective – and raises questions about the independence of the individuals. If the directors have been there for twice as long as the average director, can they still be considered independent board members?

Our data viz app tries to answer the following questions:

• Do the board members and directors of a company that I want to invest in have any influence on its financial performance? If so, what are all the insights/information I can get in this regard with this app and what do I need to research further?

• How are the board members, directors, or chairman of the company I am interested in connected with others, and what kind of influence that might have in the company? How would that affect the governance and sustainability of the company? What insights can I get from those connections that require further research?

• Who are the influencers for the company I am interested in compared to others in the same sector and or countries?

• What is all the ESG information this app can provide for the company directors? How well represented are females in this space? How does the company compare to its peers on these metrics (similar market cap, country, sector)?

• How to target initiatives to improve Gender representation – what industries and/or countries and/or market cap has a weaker record?

• Are there countries that companies seek out for establishing headquarters in because they allow greater Board influence?

• Are there specific blocks of countries that share significant connections across all sectors? Can we find other data that will show that this indicates government involvement?

## IV. DATA LIMITATIONS

A few data constraints were observed, and the tool could provide better insights if those limitations can be addressed before deploying similar dashboards.

1. Influence scores are derived by an algorithm that incorporates all the various Director attributes. In the example in Fig. 12, John Malone is the Chairman of two companies, in one of them he has very high INF scores, as well as the one where he is a founder. One possible reason for the lower score in the second company he is Chairman of could be a significantly smaller number of shares. That information is not available in the data provided by Free Float LLC and would prompt further research.

| Director | Company Name | Age | Gender | Tenure | FOUNDER | CEO | CHAIRMAN | FOUNDER_FIRM | OWNERSHIP_CATEGORY | inf_2018 | inf_2019 | inf_2020 | inf_2021 | inf_2022 |
|---|---|---|---|---|---|---|---|---|---|---|---|---|---|---|
| John Malone | LIBERTY BROADBAND CORPORATION | 81 | Male | 8 | | | T | No | Controlled | 65.83 | 65.83 | 65.83 | 62.79 | 62.79 |
| John Malone | LIBERTY GLOBAL PLC | 81 | Male | 17 | | | T | No | Controlled | | | 9.36 | 9.79 | 10.07 |
| John Malone | QURATE RETAIL, INC. | 81 | Male | 28 | T | | | Yes | Principal Shareholder | 60.5 | 59.2 | 59.2 | 59.36 | 60.6 |
| John Malone | WARNER BROS. DISCOVERY, INC. | 81 | Male | 14 | | | | No | Principal Shareholder | 7.42 | 6.91 | 7.24 | 6.36 | 6.36 |

Fig. 12. *Influence Score Constraint*

2. Director Information in the Director Independence data set (Fig. 13) seemed to be incomplete or just not in sync with the Board Connection Edges data set (Fig. 14). Directors were listed in the Edges table many more times (many more companies). The networking graphs would be more complete if we had information on all directors existing in the Edges table.

| Director ID | Director Full Name | Company Name |
|---|---|---|
| 19063 | Shirley Jackson | FEDEX CORPORATION |
| 19063 | Shirley Jackson | KYNDRYL HOLDINGS, INC. |

Fig. 13. *Directors Independence Table*

| ISSUER_NAME | Director |
|---|---|
| FEDEX CORPORATION | Shirley Jackson |
| INTERNATIONAL BUSINESS MACHINES CORPORATION | Shirley Jackson |
| KEYCORP | Shirley Jackson |
| KYNDRYL HOLDINGS, INC. | Shirley Jackson |
| MARATHON OIL CORPORATION | Shirley Jackson |
| MEDTRONIC PUBLIC LIMITED COMPANY | Shirley Jackson |
| PUBLIC SERVICE ENTERPRISE GROUP INCORPORATED | Shirley Jackson |
| SEALED AIR CORPORATION | Shirley Jackson |
| UNITED STATES STEEL CORPORATION | Shirley Jackson |

Fig. 14. *Board Connection Edges Table*

3. We did not have any information on how Directors at the Connection Edge table were assigned as either Node 1 or Node 2, except that those connections are professional ones (overlap of time worked in companies - Fig. 15). It is important to define the meaning of this order to allow revealing more specific information and importance to the connections. Fracassi (2017) presents definitions for each set of connections such as current and past employment, education, and others (memberships in clubs, organizations, charities, etc.).

If extra data could be gathered with other kinds of connections like the ones defined by Fracassi, the tool would provide further insights.

Fig. 15. *Undefined Rules for N1 and N2*

V.  FUTURE DEVELOPMENT

Once the data constraints are addressed, the tool can be updated to provide more complete node tracing. Something similar to the 'six degrees of Kevin Bacon' algorithm could be implemented, where you could select two directors and the tool would show the shortest path that connects them (Fig. 16).

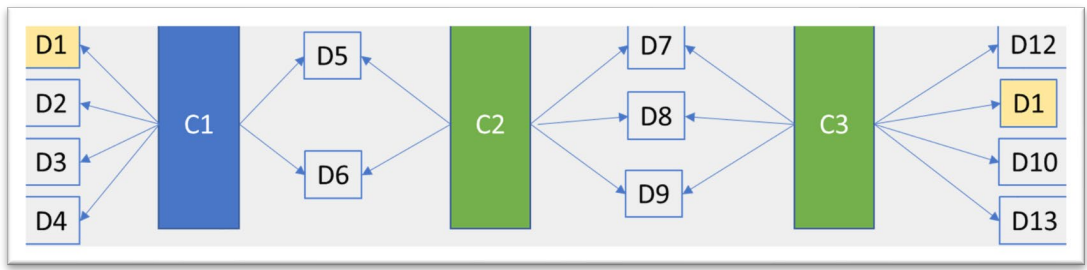

Fig. 16. *Company Network of Connections*

Since new data is mined with a certain frequency and we did not work with the full set of directors, this application involves periodic data refreshing. Moreover, the director profile contains unstructured data that has not been developed to its full extent. This data point contains valuable director information that, if analyzed, could reveal further network information such as nonprofit work, previous employment at private companies, government positions, etc. Being able to incorporate this metadata for the directors could increase the information available to the end user.

If those executive social network dashboards are used alongside other finance indicators such as R&D, SG&A expenses, cash reserves, and interest coverage ratio, for example, it could close a loop for meaningful investment insights and provide a plan for further dashboard developments.

## VI. CONCLUSION

This Data Viz App provides insights on board members and directors of a company that one would like to invest in, how they are connected, and what kind of influence they might have in the company. It can locate who are the influencers for the company one wants to invest in and compare to others in the same sector and or countries. It provides comparison among companies given several attributes such as market cap, country, sector, influencers, gender, and age of their directors and most important teases out a few questions that could only be answered with further research beyond the available data. These factors all affect a company's governance and sustainability. As such, they are useful to investors from both an investment return standpoint and as a tool for shareholder activism, as well as to the board of directors themselves for self-monitoring.

This project illustrates how the partnership of industry and academia is effective and extremely beneficial for both. The exploration of this data allowed the students to practice using a rich and complex data set and to learn advanced dashboarding techniques, even though the course is more of an introductory and intermediate one. It also sharpened their analytical reasoning and communication skills, demonstrated by their presentations to the data owner at the conclusion of the course. Inference is a sought-after skill in the data analytics/science market and this kind of project allows inference skills to flourish.

This project generated a valuable proof of concept for our industry sponsor, as they are now able to quickly visualize board member connections and influence. Their clients were impressed with the storytelling the dashboard was able to provide. When shown the charts of gender influence by country, or specific case studies about tenure or connections at certain companies, clients were not only surprised by the findings, but it generated more questions that are able to be answered by selecting/deselecting filters. Both factors feed a loop of experiential learning using real data and delivering a sought market tool that can be augmented by the company sponsoring the project.

It is important that Free Float LLC monitors the veracity of the incoming data so that the application can have real value to investors. Trying to incorporate precise information about the node connections and possibly the origin of those connections will provide even richer data variety.

## ACKNOWLEDGMENT

To Free Float LLC, who have provided such rich experiences to Northeastern University students.